\documentclass[aps,twocolumn,showpacs,showkeys]{revtex4}
\usepackage{amsmath,latexsym,amssymb}
\usepackage[dvips]{graphicx}
\usepackage{amsmath}
\usepackage{latexsym}
\usepackage{amssymb}
\usepackage{bm}

\begin{document}
\title{New Method to Reveal the Conflict Between Local Realism and Quantum Mechanics}
\author{Koji  \surname{Nagata}}
\author{Jaewook  \surname{Ahn}}
\affiliation{ Department of Physics, Korea Advanced Institute of
Science and Technology, Daejeon 305-701, Korea}

\begin{abstract}
We formulate the expectation value of the Bell-\.Zukowski operator
acting on qubit states of a two-particle Bell experiment. By using
the equivalence between a set of $N$ copies of a two-qubit
experiment and a standard two-setting Bell experiment in an
entangled $2N$-particle state, we obtain an inequality, which we may
call the Bell-\.Zukowski inequality. It determines whether the
measured correlation functions of two-particle states can be modeled
locally and realistically. In this Bell experiment of two particles,
the conflict between local realism and quantum mechanics is
discussed in conjunction with the violation of the Bell-\.Zukowski
inequality. The main point of the result is that the Bell-\.Zukowski
operator can be represented by the Bell-Mermin operator. The
threshold visibility of two-particle interference analyzed in this
scheme shows good agreement with the value to cause a violation of
the Bell-\.Zukowski inequality.
\end{abstract}


\date{\today}

\maketitle

\section{Introduction}

Bell inequalities that correlation functions satisfying local
realistic theories must obey can be violated by certain quantum
predictions, as Bell reported in 1964~\cite{bib:Bell}. Bell used the
singlet state, or EPR pairs~\cite{EPR}, to show that the correlation
functions measured in such singlet states cannot be modeled by local
realistic models. Likewise, a certain set of correlation functions
produced by quantum measurements of a quantum state contradicts
certain predictions of local realistic theories. Those states also
cannot be modeled by local realistic models. Up to now, local
realistic theories have been studied
extensively~\cite{bib:Redhead,bib:Peres,bib:lee}. Many experiments
have shown that Bell inequalities and local realistic theories are
violated~\cite{experiment1a,experiment1b,experiment2,experiment3a,experiment3b}.
Later, in a work by Fine~\cite{bib:fine}, a set of correlation
functions can be described with the property that they are
reproducible by local realistic theories for a system in two-partite
states if and only if the set of correlation functions satisfies the
complete set of (two-setting) Bell inequalities. This result is
generalized~\cite{bib:Zukowski,bib:gene} to a system described by
multipartite states in the case where two dichotomic observables are
measured per site.

In this paper, we present a method using two Bell
operators~\cite{Bellope} to refute local realistic models of a
quantum state. In order to do so, we need only a two-setting and
two-particle Bell experiment reproducible by local realistic
theories. Such a Bell experiment also reveals the conflict between
local realism and quantum mechanics in the sense that the
Bell-\.Zukowski inequality \cite{bib:Zukowski2} is violated.

Let us consider two-qubit states that, under specific settings, give
correlation functions reproducible by local realistic theories.
Imagine that $N$ copies of the states can be distributed among $2N$
parties in such a way that each pair of parties shares one copy of
the state. The parties perform a Bell-Greenberger-Horne-Zeilinger
(GHZ) $2N$-particle
experiment~\cite{bib:Zukowski,bib:gene,bib:Mermin} on their qubits.
Each of the pairs of parties uses the measurement settings noted
above. The Bell-Mermin operator~\cite{Bellope,bib:Werner2}, $B$, for
their experiment does not show any violation of local realism.
Nevertheless, one can find another Bell operator, which differs from
$B$ by a numerical factor, that does show such a violation. That is,
the original two-qubit states cannot be modeled by local realistic
models.

More specifically, the situation is as follows: A given two-setting
and two-particle Bell experiment is reproducible by local realistic
theories. However, the experimental correlation functions can
compute a violation of the Bell-\.Zukowski inequality. Therefore,
actually measured data reveal that the measured state cannot be
modeled by local realistic models. Thus, a conflict between local
realism and quantum mechanics is revealed. We can see this
phenomenon by the simple algebra presented below.

This phenomenon can occur when the system is in a mixed two-qubit
state. We analyze the threshold visibility for two-particle
interference to reveal the conflict mentioned above. It is found
that the threshold visibility agrees with the value to obtain a
violation of the Bell-\.Zukowski inequality.

\section{Bell-Mermin operator and Bell-\.Zukowski operator}

\begin{figure}[t!]
\begin{center}
\includegraphics[width=0.25\textwidth]{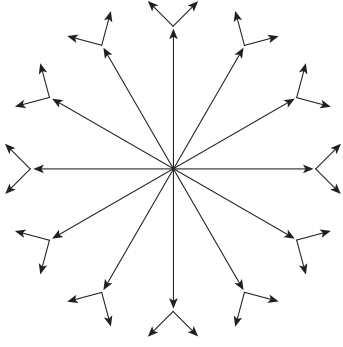}
\caption{Schematic diagram of a standard two-setting Bell experiment
in an entangled state in twelve particles with the Bell-Mermin
operator $B_{{\bf N}_{2N}}=2^{(2N-1)/2}(|\Psi^{+}_0\rangle\langle
\Psi^{+}_0|- |\Psi^{-}_0\rangle\langle \Psi^{-}_0|)$ acting on
Greenberger-Horne-Zeilinger states %
$|\Psi^{\pm}_0\rangle = (|0^{\otimes 2N}\rangle\pm|1^{\otimes
2N}\rangle)/\sqrt{2}$.} \label{standard}\end{center}
\end{figure}

Let ${\bf N}_{2N}$ be $\{1,2,\ldots,2N\}$. We consider the following
specific Bell-Mermin operator (see Eq.~(\ref{Mermin operator})):
\begin{eqnarray}
B_{{\bf N}_{2N}}=2^{(2N-1)/2}(|\Psi^{+}_0\rangle\langle \Psi^{+}_0|-
|\Psi^{-}_0\rangle\langle \Psi^{-}_0|)\label{BMineq}.
\end{eqnarray}
Here, the states $|\Psi^{\pm}_0\rangle$ are GHZ states
\cite{bib:GHZ}, i.e.,
\begin{eqnarray}
|\Psi^{\pm}_0\rangle=\frac{1}{\sqrt{2}}
(|0^{\otimes 2N}\rangle\pm|1^{\otimes 2N}\rangle).
\end{eqnarray}
An average of the Bell-Mermin operator is evaluated by using a
standard two-setting Bell experiment. See Fig.~\ref{standard}.

\begin{figure}[t!]
\begin{center}
\includegraphics[width=0.25\textwidth]{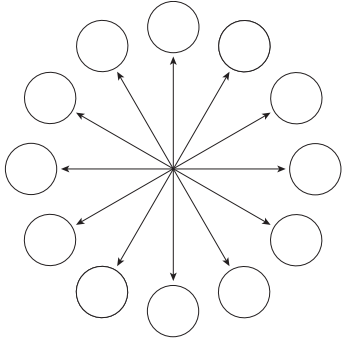}
\caption{Schematic diagram of a Bell-\.Zukowski experiment in an
entangled state of twelve particles with the Bell-\.Zukowski
operator ${Z}_{2N}=\frac{1}{2}\left(\frac{\pi}{2}\right)^{2N}
\frac{1}{2^{(2N-1)/2}}
 {B}_{{\bf N}_{2N}}$.
}\label{zukowski}\end{center}
\end{figure}

One can introduce a $2N$-partite Bell operator, which one may call
the Bell-\.Zukowski operator $Z_{2N}$, which differs from ${B}_{{\bf
N}_{2N}}$ only by a numerical factor. The Bell-\.Zukowski operator
${Z}_{2N}$ \cite{KNJA} is
\begin{eqnarray}
{Z}_{2N}=\frac{1}{2}\left(\frac{\pi}{2}\right)^{2N}
(|\Psi^+_0\rangle\langle\Psi^+_0|-
|\Psi^-_0\rangle\langle\Psi^-_0|).\label{ZukoBellope}
\end{eqnarray}
An average of the Bell-\.Zukowski operator is evaluated by using an
all-setting Bell experiment. See Fig.~\ref{zukowski}.

\begin{figure}[t!]
\begin{center}
\includegraphics[width=0.48\textwidth]{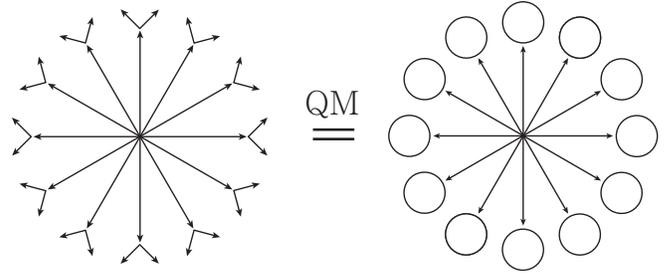}
\caption{Schematic diagram of the equivalence between a
Bell-\.Zukowski experiment and a standard two-setting Bell
experiment under the validity of quantum mechanics.}
\label{equal}\end{center}
\end{figure}

Clearly, we see that the Bell-Mermin operator given in
Eq.~(\ref{BMineq}) is connected to the Bell-\.Zukowski operator
${Z}_{2N}$ in the following relation:
\begin{eqnarray}
&&{Z}_{2N}=\frac{1}{2}\left(\frac{\pi}{2}\right)^{2N}
\frac{1}{2^{(2N-1)/2}}
 {B}_{{\bf N}_{2N}}.\label{Bellrelation}
\end{eqnarray}
One can see that the specific two-setting Bell $2N$-particle
experiment in question computes an average value of the
Bell-\.Zukowski operator $\langle{Z}_{2N}\rangle$ when an average
value of $\langle {B}_{{\bf N}_{2N}}\rangle$ is evaluated. Of
course, this argument is due to the validity of quantum mechanics.
See Fig.~\ref{equal}.

From the Bell-\.Zukowski inequality
\begin{eqnarray}%
|\langle {Z}_{2N}\rangle|\leq 1,
\end{eqnarray}%
we have a condition on the average of the Bell-Mermin operator
$\langle {B}_{{\bf N}_{2N}}\rangle$, which is written by
\begin{eqnarray}
|\langle {B}_{{\bf N}_{2N}}\rangle|\leq
2\left(\frac{2}{\pi}\right)^{2N} 2^{({2N}-1)/2}.\label{newbell}
\end{eqnarray}
Please notice that the Bell-\.Zukowski inequality $|\langle
Z_{2N}\rangle|\leq 1$ is derived under the assumption that there are
predetermined `hidden' results of the measurement for all directions
in the rotation plane for the system in a state. On the other hand,
the Bell-Mermin inequality is derived under the assumption that
there are predetermined `hidden' results of the measurement for two
directions for the system in a state. We see that a violation of the
condition in Eq.~(\ref{newbell}) implies a violation of the
Bell-\.Zukowski inequality. Our aim is to compute an expectation
value of the Bell-Mermin operator given in Eq.~(\ref{BMineq}) by
using a two-particle Bell experiment reproducible by local realistic
theories. The Bell-\.Zukowski inequality is stronger than the
standard Bell inequalities for $N\geq 2$. This is why a standard
Bell experiment reproducible by local realistic theories reveals the
conflict between local realism and quantum mechanics.

\section{Experimental situation}

We consider the following two-qubit states:
\begin{eqnarray}
\rho_{a,b}=V|\psi\rangle\langle\psi|+(1-V)\rho_{\rm noise} ~
(0\leq V\leq 1),\label{qubit}
\end{eqnarray}
where $|\psi\rangle$ is a Bell state defined as
\begin{eqnarray}%
|\psi\rangle=\frac{1}{\sqrt{2}}(|+^a;+^b\rangle-i|-^a;-^b\rangle).
\end{eqnarray}%
$\rho_{\rm noise} = \frac{1}{4} \openone$ is the random noise
admixture. The value of $V$ can be interpreted as the reduction
factor of the interferometric contrast observed in the two-particle
correlation experiment. The states $| \pm^k \rangle$ are eigenstates
of the $z$-component of the Pauli observable, $\sigma_z^k$, for the
$k$th observer. Here, $a$ and $b$ are the labels of the parties (say
Alice and Bob). Then, we have $ {\rm
tr}[\rho_{a,b}\sigma^a_x\sigma^b_x]=0, {\rm
tr}[\rho_{a,b}\sigma^a_y\sigma^b_y]=0, {\rm
tr}[\rho_{a,b}\sigma^a_x\sigma^b_y]=V,$ and ${\rm
tr}[\rho_{a,b}\sigma^a_y\sigma^b_x]=V.$

Here, $\sigma^k_x$ and $\sigma^k_y$ are Pauli-spin operators for the
$x$-component and for the $y$-component, respectively. This set of
experimental correlation functions is described with the property
that they are reproducible by local realistic theories. See the
following relations along with the arguments in Ref.~10:
\begin{widetext}
\begin{eqnarray}
&&|{\rm tr}[\rho_{a,b}\sigma^a_x\sigma^b_x]
-{\rm tr}[\rho_{a,b}\sigma^a_y\sigma^b_y]
+{\rm tr}[\rho_{a,b}\sigma^a_x\sigma^b_y]
+{\rm tr}[\rho_{a,b}\sigma^a_y\sigma^b_x]|=2V \leq 2,\nonumber\\
&&|{\rm tr}[\rho_{a,b}\sigma^a_x\sigma^b_x]
+{\rm tr}[\rho_{a,b}\sigma^a_y\sigma^b_y]
-{\rm tr}[\rho_{a,b}\sigma^a_x\sigma^b_y]
+{\rm tr}[\rho_{a,b}\sigma^a_y\sigma^b_x]|=0 \leq 2,\nonumber\\
&&|{\rm tr}[\rho_{a,b}\sigma^a_x\sigma^b_x]
+{\rm tr}[\rho_{a,b}\sigma^a_y\sigma^b_y]
+{\rm tr}[\rho_{a,b}\sigma^a_x\sigma^b_y]
-{\rm tr}[\rho_{a,b}\sigma^a_y\sigma^b_x]|=0\leq 2,\nonumber\\
&&|{\rm tr}[\rho_{a,b}\sigma^a_x\sigma^b_x]
-{\rm tr}[\rho_{a,b}\sigma^a_y\sigma^b_y]
-{\rm tr}[\rho_{a,b}\sigma^a_x\sigma^b_y]
-{\rm tr}[\rho_{a,b}\sigma^a_y\sigma^b_x]|=2V \leq 2.\label{LHV}
\end{eqnarray}
\end{widetext}

In the following section, we will use this kind of experimental
situation. Those experimental correlation functions
can compute a violation of the Bell-\.Zukowski inequality. In order
to do so, we shall compute an expectation value of the
Bell-Mermin operator in the
next section.

\section{Conflict between local realism and
quantum mechanics}

\begin{figure}[t!]
\begin{center}
\includegraphics[width=0.48\textwidth]{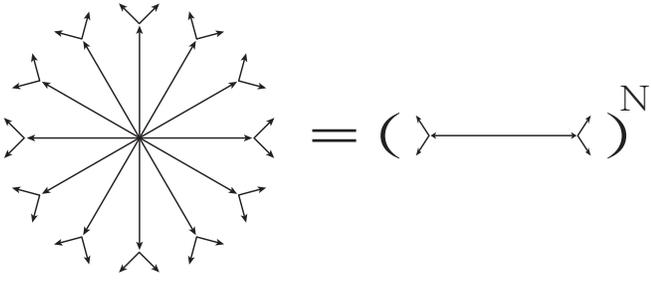}
\caption{Schematic diagram of $N$ copies of two-qubit experiments
which are equivalent to a standard two-setting Bell experiment in an
entangled $2N$-particle state.} \label{experiment}\end{center}
\end{figure}

Imagine that $N$ copies of the states introduced in the preceding
section can be distributed among $2N$ parties in such a way that
each pair of parties shares one copy of the state
\begin{eqnarray}
\rho^{\otimes N}=\underbrace{\rho_{1,2}\otimes\rho_{3,4}\otimes
\cdots\otimes\rho_{N-1,N}}_{N}.
\end{eqnarray}
Suppose that spatially separated $2N$ observers perform measurements
on each of $2N$ particles. The decision processes for choosing
measurement observables are space-like separated. It can be regarded
as a standard two-setting Bell  experiment in an entangled state in
$2N$ particles. See Fig.~\ref{experiment}.

We assume that a two-orthogonal-setting Bell-GHZ $2N$-particle
correlation experiment is
performed. We choose measurement observables such that
\begin{eqnarray}
A_k=\sigma^k_x,
A'_k=\sigma^k_y.
\label{setting}
\end{eqnarray}
Namely, each of the pairs of parties uses measurement settings such
that they can check the condition in Eq.~(\ref{LHV}). Therefore, it
should be that given $2^{2N}$ correlation functions are described
with the property that they are reproducible by local realistic
theories. The Bell-Mermin operators ${B}_{{\bf N}_{2N}}$ and
${B}'_{{\bf N}_{2N}}$ do not show any violation of local realism as
shown below.

Let $f(x,y)$ denote the function
\begin{eqnarray}
\frac{1}{\sqrt{2}}e^{-i\pi/4}(x+iy), x,y\in {\bf R}.
\end{eqnarray}%
$f(x,y)$ is invertible as \begin{eqnarray}%
 x=\Re f-\Im f, y=\Re
f+\Im f .%
\end{eqnarray}%
The Bell-Mermin operators ${B}_{{\bf N}_{2N}}$ and ${B}'_{{\bf
N}_{2N}}$ are defined by \cite{bib:Mermin,bib:Werner2} %
\begin{eqnarray} f({B}_{{\bf N}_{2N}},{B}'_{{\bf N}_{2N}})=\prod_{k=1}^{2N}
f(A_k,A'_k) . %
\end{eqnarray}%
The Bell-Mermin inequality can be expressed as \cite{bib:Werner2}
\begin{eqnarray}
|\langle{B}_{{\bf N}_{2N}}\rangle|\leq 1, ~~|\langle{B}'_{{\bf N}_{2N}}
\rangle|\leq 1.\label{MKin}
\end{eqnarray}
We also define ${B}_{\alpha}$ for any subset $\alpha\subset{\bf N}_{2N}$ by
\begin{eqnarray}
f({B}_{\alpha},{B}'_{\alpha})=\prod_{k\in \alpha}
f(A_k,A'_k).
\end{eqnarray}
It is easy to see that, when $\alpha, \beta(\subset{\bf N}_{2N})$ are disjoint,
\begin{eqnarray}
f({B}_{\alpha\cup \beta},
{B}'_{\alpha\cup \beta})=f({B}_{\alpha},{B}'_{\alpha})\otimes
f({B}_{\beta},
{B}'_{\beta}),
\end{eqnarray}
which leads to following equations:
\begin{eqnarray}
{B}_{\alpha\cup \beta}&=&
(1/2){B}_{\alpha}\otimes({B}_{\beta}
+{B}'_{\beta})
+
(1/2){B}'_{\alpha}\otimes({B}_{\beta}-
{B}'_{\beta}),\nonumber\\
{B}'_{\alpha\cup \beta}&=&
(1/2){B}'_{\alpha}\otimes({B}'_{\beta}
+{B}_{\beta})
+
(1/2){B}_{\alpha}\otimes({B}'_{\beta}-
{B}_{\beta}).\nonumber\\
\label{GB}
\end{eqnarray}
In specific operators $A_k$ and $A'_k$ given in Eq.~(\ref{setting}),
where %
\begin{eqnarray}
\sigma^k_x=|+^k\rangle\langle-^k|+|-^k\rangle\langle+^k|%
\end{eqnarray}%
and %
\begin{eqnarray}
\sigma^k_y=-i|+^k\rangle\langle-^k|+i|-^k\rangle\langle+^k|, %
\end{eqnarray}
we have (cf. \cite{bib:scarani})
\begin{eqnarray}
&&f(A_k,A'_k)=(e^{-i\frac{\pi}{4}}/\sqrt{2})
(\sigma_x^k+i\sigma^k_y)\nonumber\\
&&=e^{-i\frac{\pi}{4}}\sqrt{2}|+^k\rangle\langle-^k|
\end{eqnarray}
and
\begin{eqnarray}
&&f({B}_{{\bf N}_{2N}},{B}'_{{\bf N}_{2N}})=
\prod_{k=1}^{2N} f(A_k,A'_k)\nonumber\\
&&=e^{-i\frac{2N\pi}{4}} 2^{N}\prod_{k=1}^{2N} |+^k\rangle\langle-^k|
\nonumber\\
&&=e^{-i\frac{2N\pi}{4}} 2^{N}|+^{\otimes 2N}\rangle\langle-^{\otimes 2N}|.
\end{eqnarray}
Hence, we obtain
\begin{eqnarray}
{B}_{{\bf N}_{2N}}
&=&
2^{N}\big\{(1/2)(e^{-i\frac{2N\pi}{4}}
|+^{\otimes 2N}\rangle\langle-^{\otimes 2N}|
+H.c.)\nonumber\\
&&-(-i/2)
(e^{-i\frac{2N\pi}{4}}
|+^{\otimes 2N}\rangle\langle-^{\otimes 2N}|
 -H.c.)
\big\}
\nonumber\\
&&=2^{\frac{2N-1}{2}}(e^{-i\frac{(2N-1)\pi}{4}}|+^{\otimes 2N}\rangle
\langle-^{\otimes 2N}|
+ H.c.)\nonumber\\
&&=2^{(2N-1)/2}(|\Psi^{+}_0\rangle\langle \Psi^{+}_0|-
|\Psi^{-}_0\rangle\langle \Psi^{-}_0|),\label{Mermin operator}
\end{eqnarray}
where %
\begin{eqnarray}
e^{-i\frac{(2N-1)\pi}{4}}|+^{\otimes 2N}\rangle= |1^{\otimes
2N}\rangle.
\end{eqnarray}

Measurements on each of $2N$ particles enable them to obtain
$2^{2N}$ correlation functions. Thus, they get an average value of
the specific Bell-Mermin operator given in Eq.~(\ref{BMineq}).
According to Eq.~(\ref{GB}), we obtain
\begin{eqnarray}
\langle {B}_{{\bf N}_{2N}} \rangle=
\langle {B}'_{{\bf N}_{2N}} \rangle=
\prod_{i=2}^{N}\langle {B}_{\{i-1,i\}} \rangle=
V^{N}(\leq 1).
\end{eqnarray}
Clearly, the Bell-Mermin operators ${B}_{{\bf N}_{2N}}$ and
${B}'_{{\bf N}_{2N}}$ for their experiment do not show any violation
of local realism as we have mentioned above.

Nevertheless, when $N\geq 2$ and $V$ is given by
\begin{equation}
\left(2\left(\frac{2}{\pi}\right)^{2N} 2^{(2N-1)/2}\right)^{1/N}<V \leq 1,
\label{mainresult}
\end{equation}
we have a violation of the condition in Eq.~(\ref{newbell}), i.e.,
one can compute a violation of the Bell-\.Zukowski inequality
$|\langle {Z}_{2N}\rangle|\leq 1$ that is, the measured two-qubit
state cannot be modeled by local realistic models. The condition in
Eq.~(\ref{mainresult}) says that the threshold visibility decreases
when the number of copies, $N$, increases. In an extreme situation,
when $N\rightarrow \infty$, we
have the desired condition %
\begin{eqnarray}
V>2(2/\pi)^2 \label{Zukowskimin}%
\end{eqnarray}%
to show the conflict in question. This agrees with the value to get
a violation of the Bell-\.Zukowski inequality. It is worth noting
that the condition in Eq.~(\ref{Zukowskimin}) gives $V > 0.81$,
which does not seem to conflict with the condition in
Eq.~(\ref{mainresult}).

The given example using two-qubit states reveals the violation of
the Bell-\.Zukowski inequality. The interesting point is that all
the information to get the violation of the Bell-\.Zukowski
inequality can be obtained only by a two-setting and two-particle
Bell experiment reproducible by local realistic theories.

\section{SUMMARY}

In summary, we have shown that the Bell-\.Zukowski operator can be
represented by the Bell-Mermin operator. This fact provides a means
to check whether a quantum state can be modeled by local realistic
models, i.e., if the conflict between local realism and quantum
mechanics occurs. Our argument relies only on a two-setting and
two-particle Bell experiment reproducible by local realistic
theories. Given a two-setting and two-particle Bell experiment
reproducible by local realistic theories, one can compute a
violation of the Bell-\.Zukowski inequality. Measured data, thus,
indicate that the measured state cannot be modeled by local
realistic models. Thus, the conflict between local realism and
quantum mechanics is revealed. This phenomenon can occur when the
system is in a mixed state. We also analyzed the threshold
visibility for two-particle interference in order to bring about the
phenomenon. The threshold visibility agrees well with the value to
obtain a violation of the Bell-\.Zukowski inequality.


\section*{ACKNOWLEDGEMENTS}
This work was supported by the Frontier Basic Research Programs at
Korea Advanced Institute of Science and Technology. K.~N. is
supported by a BK21 research grant.



\newpage

\newpage

\newpage

\newpage

\end{document}